\documentstyle[12pt,epsfig]{article}
\def\lbldef#1#2{\expandafter\gdef\csname #1\endcsname {#2}}

\def\href#1#2{#2}  
\begin{document}
\baselineskip=15.5pt
\pagestyle{plain}
\setcounter{page}{1}
%\renewcommand{\thefootnote}{\fnsymbol{footnote}}
%--------+---------+---------+---------+---------+---------+---------+
%Title page

\begin{titlepage}

\begin{flushright}
CERN-TH/2000-027\\
hep-th/0001166
\end{flushright}
\vspace{10 mm}

\begin{center}
{\Large A Note on Solitons in Brane Worlds}

\vspace{5mm}

\end{center}

\vspace{5 mm}

\begin{center}
{\large Donam Youm\footnote{E-mail: Donam.Youm@cern.ch}}

\vspace{3mm}

Theory Division, CERN, CH-1211, Geneva 23, Switzerland

\end{center}

\vspace{1cm}

\begin{center}
{\large Abstract}
\end{center}

\noindent

We obtain the zero mode effective action for gravitating objects in the 
bulk of dilatonic domain walls.  Without additional fields included in the 
bulk action, the zero mode effective action reproduces the action in one 
lower dimensions obtained through the ordinary Kaluza-Klein (KK) 
compactification, only when the transverse (to the domain wall) component 
of the bulk metric does not have non-trivial term depending on the domain 
wall worldvolume coordinates.  With additional fields included in the bulk 
action, non-trivial dependence of the transverse metric component on 
the domain wall worldvolume coordinates appears to be essential in 
reproducing the lower-dimensional action obtained via the ordinary KK 
compactification.  We find, in particular, that the effective action for the 
charged $(p+1)$-brane in the domain wall bulk reproduces the action for the 
$p$-brane in one lower dimensions.

\vspace{1cm}
\begin{flushleft}
CERN-TH/2000-027\\
January, 2000
\end{flushleft}
\end{titlepage}
\newpage

\section{Introduction}

Recently, phenomenolgists have actively considered the possibility that 
the existence of additional compact spatial dimensions may account 
for the large hierarchy between the electroweak scale and the Planck 
scale, the so-called hierarchy problem in particle physics.  In this 
scenario, our four-dimensional world is confined within the worldvolume 
of a three-brane, within which the fields of Standard Model are contained.  
The earlier proposal \cite{ad1,ad2} relies on the large enough volume of 
the compact extra space for solving the hierarchy problem.  More compelling 
scenario proposed by Randall and Sundrum (RS) \cite{ran1,ran2,ran3} 
assumes that the spacetime is non-factorizable, in contrast to the 
conventional view of the Kaluza-Klein (KK) theory that the spacetime is the 
direct product of the four-dimensional spacetime and the compact extra space.  
Such a point of view on spacetime was also previously taken 
\cite{ncc1,ncc2,ncc3,ncc4,ncc5,ncc6,ncc7,ncc8} as an alternative to the 
compact compactification, namely as a mechanism to trap matter within the 
four-dimensional hypersurface without having to assume that the extra space 
is compact.  However, what is new in the RS model is that it is 
not just matter but also gravity that is effectively trapped within the 
hypersurface, i.e. the three-brane or the five-dimensional domain wall.  
The exponential fall-off (as one moves away from the brane) of the warp 
factor in the metric of the non-factorizable spacetime accounts for the 
large hierarchy between the electroweak and the Planck scales in our 
four-dimensional world, which is assumed to be located away from the wall.  
Gravity in our four-dimensional world is relatively weak also because the 
wave function for the graviton zero mode, which is localized around the 
wall, falls off as one moves away from the wall.

Since gravity is shown to be effectively compactified to one lower 
dimensions (even when the extra spatial dimension is infinite) in the 
background of the RS type domain walls \cite{ran2}, it is of interest to 
study various gravitating objects in such background.  (Some of the 
previous works on the related subject are Refs. \cite{cg,chr,ehm,gs,ehm2}.) 
In our previous works \cite{youm1,youm2,youm3}, we attempted to understand 
charged branes in the bulk background of the RS type domain wall.  It turns 
out that the domain wall bulk background is so restrictive about the 
possible gravitating objects that non-dilatonic domain wall 
bulk background in general does not allow charged branes.  One of ways 
to get around this difficulty is to allow the cosmological 
constant term in the bulk action to have the dilaton factor.  
Fortunately, it is observed \cite{youm1} that even the dilatonic domain 
wall background effectively compactifies gravity, if the dilaton coupling 
parameter in the cosmological constant term is small enough.  The 
warp factor in the metric of the dilatonic domain wall with small 
enough dilaton coupling parameter also decreases as one moves away 
from the wall within the finite allowed coordinate interval around the 
wall as a power-law, instead of exponentially within the infinite 
allowed coordinate interval around the wall just like non-dilatonic domain 
wall of the RS model \cite{ran1,ran2,ran3}, and becomes zero at the end of 
the allowed finite coordinate interval.  So, one can also use such dilatonic 
domain walls for tackling the hierarchy problem. 

It is realized \cite{youm2} that charged $p$-branes, as observed in one 
lower dimensions, should rather be regarded as charged $(p+1)$-branes in 
the bulk of domain wall, because charged $p$-branes in the bulk of domain 
wall background are not effectively compactified to the charged $p$-branes 
in one lower dimensions on the hypersurface of the wall.  We studied 
\cite{youm3} the dynamics of probes in the background of such charged 
branes for the purpose of understanding the spacetime properties of charged 
branes in the bulk background of the domain walls.  In this paper, we 
check whether such charged $(p+1)$-branes in the domain wall bulk effectively 
describe the corresponding charged $p$-branes in one lower dimensions 
or describe different physics in one lower dimensions by obtaining the 
effective action (in one lower dimensions) for such charged $(p+1)$-branes 
in the bulk of the domain walls.  We find that the effective action has 
exactly the same form as the action for the $p$-brane in one lower dimensions 
that is obtained from the action for the $(p+1)$-brane through the ordinary 
KK compactification on $S^1$.  

The paper is organized as follows.  In section 2, we discuss dilatonic 
domain wall solution that we studied in our previous works.  In sections 3, 
we obtain the effective action for the charged branes in the bulk of the 
dilatonic domain wall.

\section{Dilatonic Domain Wall Solution}

We begin by discussing the $D$-dimensional extreme dilatonic domain wall 
solution studied in Ref. \cite{youm3}.   Generally, the total action for the 
RS type model is the sum of the $D$-dimensional action $S_{\rm bulk}$ 
for the fields in the bulk of the domain wall and the $(D-1)$-dimensional 
action $S_{\rm DW}$ on the domain wall worldvolume:
\begin{equation}
S=S_{\rm bulk}+S_{\rm DW}.
\label{ddact}
\end{equation}
The bulk action contains the following action for the dilatonic domain 
wall solution:
\begin{equation}
S_{\rm bulk}\supset {1\over{2\kappa^2_D}}\int d^Dx\sqrt{-G}\left[
{\cal R}_G-{4\over{D-2}}\partial_M\phi\partial^M\phi+e^{-2a\phi}
\Lambda\right],
\label{ddwact}
\end{equation}
and the $(D-1)$-dimensional action contains the following worldvolume 
action $S_{\sigma}$ for the dilatonic domain wall:
\begin{equation}
S_{\rm DW}\supset S_{\sigma}=-\sigma_{\rm DW}\int d^{D-1}x\sqrt{-\gamma}
e^{-a\phi}.
\label{wvddwact}
\end{equation}
Here, $\sigma_{\rm DW}$ is the tension or the energy density of the domain 
wall and $\gamma$ is the determinant of the induced metric $\gamma_{\mu\nu}=
\partial_{\mu}X^M\partial_{\nu}X^NG_{MN}$ on the domain wall worldvolume, 
where $M,N=0,1,...,D-1$ and $\mu,\nu=0,1,...,D-2$.

The domain wall solution has the following form:
\begin{equation}
G_{MN}dx^Mdx^N={\cal W}\eta_{\mu\nu}dx^{\mu}dx^{\nu}+dy^2,\ \ \ \ \ \ 
e^{2\phi}={\cal W}^{{(D-2)^2a}\over{4}},
\label{dddw}
\end{equation}
where the warp factor ${\cal W}$ is given by
\begin{eqnarray}
{\cal W}(y)&=&\left(1+{{2(D-1)+(D-2)\Delta}\over{(D-2)\Delta}}Q|y|
\right)^{4\over{2(D-1)+(D-2)\Delta}}
\cr
&=&\left(1+{{a^2(D-2)^2}\over{a^2(D-2)^2-4(D-1)}}Q|y|
\right)^{8\over{a^2(D-2)^2}},
\cr
\Delta&=&{{(D-2)a^2}\over 2}-{{2(D-1)}\over{D-2}},
\label{warpfactor}
\end{eqnarray}
for $a\neq 0$, and
\begin{equation}
{\cal W}(y)=\exp\left(-{{2Q}\over{D-1}}|y|\right),
\label{warpa0}
\end{equation}
for $a=0$.  The domain wall solution (\ref{dddw}) solves the equations of 
the motion of the combined actions (\ref{ddwact}) and (\ref{wvddwact}), 
provided the bulk cosmological constant $\Lambda$ and the domain wall 
tension $\sigma_{\rm DW}$ in the action are related to the parameter $Q$ in 
the solution as
\begin{equation}
\Lambda=-{{2Q^2}\over\Delta},\ \ \ \ \ \ \ \ \ 
\sigma_{\rm DW}={4\over{|\Delta|}}{Q\over{\kappa^2_D}}.
\label{relqcos}
\end{equation}
In this paper, we assume that $Q>0$, so that the tension $\sigma_{\rm DW}$ of 
the wall is positive.

One can bring the domain wall metric (\ref{dddw}) to the conformally flat 
form by applying the following coordinate transformation:
\begin{eqnarray}
z&=&{\rm sgn}(y){{a^2(D-2)^2-4(D-1)}\over{a^2(D-2)^2-4}}Q^{-1}
\left[\left(1+{{a^2(D-2)^2}\over{a^2(D-2)^2-4(D-1)}}Q|y|
\right)^{{a^2(D-2)^2-4}\over{a^2(D-2)^2}}-1\right]
\cr
&=&{\rm sgn}(y){\Delta\over{\Delta+2}}Q^{-1}\left[\left(1+
{{2(D-1)+(D-2)\Delta}\over{(D-2)\Delta}}Q|y|\right)^{{(D-2)(\Delta+2)}\over
{2(D-1)+(D-2)\Delta}}-1\right],
\label{crdtrtr1}
\end{eqnarray}
for $a\neq 0$, and
\begin{equation}
z={\rm sgn}(y){{D-1}\over Q}\left[\exp\left({Q\over{D-1}}|y|\right)-1\right],
\label{crdtrtr2}
\end{equation}
for $a=0$.  The resulting domain wall metric has the following form:
\begin{equation}
G_{MN}dx^Mdx^N={\cal C}\left[\eta_{\mu\nu}dx^{\mu}dx^{\nu}
+dz^2\right],\ \ \ \ \ 
e^{2\phi}={\cal C}^{{(D-2)^2a}\over 4},
\label{cfddmet}
\end{equation}
where the conformal factor ${\cal C}$ is given by
\begin{eqnarray}
{\cal C}(z)&=&\left(1+{{\Delta+2}\over{\Delta}}Q|z|
\right)^{4\over{(D-2)(\Delta+2)}}
\cr
&=&\left(1+{{(D-2)^2a^2-4}\over{(D-2)^2a^2-4(D-1)}}Q|z|
\right)^{8\over{(D-2)^2a^2-4}}.
\label{ddcf}
\end{eqnarray}

In particular, the domain wall solution of the RS model \cite{ran1,ran2,ran3} 
corresponds to the $(D,a)=(5,0)$ case of the above general solution.  The 
parameter $Q$ in the above metric is related to the parameter $k$ in the 
RS domain wall metric as $Q=4k$.  Note, also the difference in the definition 
of the cosmological constant from that of Refs. \cite{ran1,ran2,ran3}, where 
the gravitational constant does not multiply the cosmological constant term 
and there is a negative sign.  So, in order to relate the cosmological 
constant $\Lambda$ in Eq. (\ref{relqcos}) of the action (\ref{ddact}) to the 
cosmological constant of Refs. \cite{ran1,ran2,ran3}, one has to multiply 
$\Lambda$ in Eq. (\ref{relqcos}) by $-1/(2\kappa^2_5)$, where $\kappa^2_5=
1/(4M^3)$ in the notation of Refs. \cite{ran1,ran2,ran3}.

\section{Zero Mode Effective Action}

In this section, we study the $(D-1)$-dimensional effective action for the 
RS type model obtained from the total action $S_{\rm bulk}+S_{\rm DW}$ 
by integrating over the extra space coordinate.  We shall consider only 
the zero modes of the fields, in part because the harmonic functions for 
branes in the domain wall bulk under consideration, i.e., those studied 
in Ref. \cite{youm3}, are independent of the extra space coordinate.

In Ref. \cite{ran2}, it is shown that gravity in one lower dimensions is 
reproduced by allowing the following form of the small perturbation of the 
domain wall metric (\ref{cfddmet}):
\begin{equation}
G_{MN}dx^Mdx^N={\cal C}\left[(\eta_{\mu\nu}+h_{\mu\nu})
dx^{\mu}dx^{\nu}+dz^2\right],
\label{metlngpurt}
\end{equation}
where the metric perturbation $h_{\mu\nu}$ satisfies the transverse 
traceless gauge condition $h^{\mu}_{\ \mu}=0=\partial^{\mu}h_{\mu\nu}$.  
However, in this paper, we consider the case where there is an additional 
perturbation, depending on the domain wall worldvolume coordinates 
$x^{\mu}$, in the extra space direction, i.e. the $y$ or the $z$ direction. 
This is motivated by the fact that the extra space component of the metric 
of charged branes in the bulk of the domain wall generally depends on the 
worldvolume coordinates of the domain wall.  So, we consider the following 
form of the $D$-dimensional metric:
\begin{equation}
G_{MN}dx^Mdx^N={\cal C}\left[g_{\mu\nu}dx^{\mu}dx^{\nu}+h^2dz^2\right],
\label{pertmet}
\end{equation}
where ${\cal C}$ is given by Eq. (\ref{ddcf}), and $g_{\mu\nu}$ and 
$h$ are zero modes, i.e., depend on the $(D-1)$-dimensional coordinates 
$x^{\mu}$, only.  
Note, the term $h^2$ in the above metric does not have anything to do with 
the radion \cite{ran1,gw1,gw2}, which determines the scale of the distance 
between the visible and the hidden domain walls of the first RS model 
\cite{ran1}, because we are considering only one domain wall, called 
the TeV brane, and assuming the extra space dimension to be infinite just 
like the second RS model \cite{ran2,ran3}.  

First, we consider the case where the total action is given by the actions 
for the dilatonic domain wall, only, i.e., the sum of the actions 
(\ref{ddwact}) and (\ref{wvddwact}).  For the Ans\" atze for the fields, we 
use Eq. (\ref{pertmet}) for the metric $G_{MN}$ and Eq. (\ref{cfddmet}) for 
the dilaton $\phi$, where ${\cal C}$ is given by Eq. (\ref{ddcf}).  Then, 
the total action reduces to the following form:
\begin{eqnarray}
S&=&{1\over{2\kappa^2_D}}\int d^Dx\sqrt{-G}\left[{\cal R}_G-{4\over{D-2}}
\partial_M\phi\partial^M\phi+e^{-2a\phi}\Lambda\right]
-\sigma_{\rm DW}\int d^{D-1}x\sqrt{-\gamma}e^{-a\phi}
\cr
&=&{1\over{2\kappa^2_D}}\int d^{D-1}xdz\sqrt{-g}\varpi^{2\over{\Delta+2}}
\left[h{\cal R}_g-8{{D-1}\over{D-2}}{Q\over\Delta}h^{-1}\left\{\delta(z)
-{{DQ}\over{4(D-1)}}\varpi^{-2}\right\}\right.
\cr
& &\ \ \ \ \ \ \ \ \ \ \ \ \ \ \ \ \ \ \ \ \ \ \ \ \ \ \ \ \ \ \ \ \ \ \ \ \ 
-\left.{{2Q^2}\over\Delta}\varpi^{-2}h\right]-
{4\over{|\Delta|}}{Q\over{\kappa^2_D}}\int d^{D-1}x\sqrt{-g}
\cr
&=&{1\over{2\kappa^2_D}}\int d^{D-1}x\sqrt{-g}\left[-{{2\Delta Q^{-1}}\over
{\Delta+4}}h{\cal R}_g-{{4Q}\over\Delta}(h^{-1}+h-2)\right],
\label{compact}
\end{eqnarray}
where $\varpi\equiv 1+{{\Delta+2}\over\Delta}Q|z|$.  Since the domain 
wall is located at $z=0$, the fields in the domain wall worldvolume action 
$S_{\sigma}$ take the forms $e^{-a\phi}=1$ and $\gamma_{\mu\nu}=g_{\mu\nu}$ 
(in the static gauge of the worldvolume action).  Note, in the last equality, 
we integrated over all the possible values of the extra space coordinate $z$.  
Namely, ${\Delta\over{\Delta+2}}Q^{-1}\leq z\leq -{\Delta\over{\Delta+2}}
Q^{-1}$ for the $-2<\Delta<0$ case, and $-\infty<z<\infty$ for the $\Delta<
-2$ case (Cf. see Eq. (\ref{ddcf})).  When $\Delta>0$, integration over all 
the possible values of $z$, i.e. $-\infty<z<\infty$, will make the Einstein 
term in the above action diverge, meaning that gravity is not effectively 
compactified.  So, in the worldvolume action $S_{\sigma}$, we have taken 
$|\Delta|=-\Delta$.  

When the dilatonic domain wall metric (\ref{cfddmet}) does not have a 
perturbation in the extra space direction, i.e., when the $D$-dimensional 
metric is of the form Eq. (\ref{pertmet}) with $h=1$, the effective action 
(\ref{compact}) reduces to the action for the $(D-1)$-dimensional general 
relativity with the gravitational constant given by
\begin{equation}
\kappa^2_{D-1}=-{{\Delta+4}\over{2\Delta}}Q\kappa^2_D,
\label{gravcnst}
\end{equation}
as can be seen from the last line of Eq. (\ref{compact}).  
Note, $\Delta$, defined in Eq. (\ref{warpfactor}), is always greater 
than $-4$ for $D>4$, so $\kappa^2_{D-1}>0$ if $\Delta<0$ and $Q>0$.  Note, 
it is essential that $\Delta$ should be negative in order for the Einstein's 
gravity in one lower dimensions to be reproduced.  Namely, if $\Delta$ had 
been positive, ($i$) the Einstein term would have diverged, ($ii$) no 
cancellation of the extra terms in the action (\ref{compact}) would have 
occurred because of the contribution from the worldvolume action $S_{\sigma}$ 
with the opposite sign (i.e., $|\Delta|=+\Delta$ for $\Delta>0$), and ($iii$) 
the gravitational constant $\kappa^2_{D-1}$ in Eq. (\ref{gravcnst}) would not 
have been positive with $Q>0$.  This is in accordance with the result of our 
previous work \cite{youm1} that gravity cannot be trapped within the domain 
wall if $\Delta$ is positive.  So, any $D$-dimensional gravitating objects 
in the dilatonic domain wall background with the bulk action given by Eq. 
(\ref{ddwact}) and the metric given by Eq. (\ref{pertmet}) with $h=1$ 
effectively describe the corresponding configurations in the general 
relativity in one lower dimensions, as long as the dilaton coupling 
parameter $a$ is such that $\Delta<0$.  This confirms that the RS model can 
be extended to the dilatonic domain wall case.  

However, when the dilatonic domain wall metric (\ref{cfddmet}) has a 
perturbation in the transverse direction, i.e., when $h$ in Eq. 
(\ref{pertmet}) is a non-trivial function of the domain wall worldvolume 
coordinates $x^{\mu}$, the effective action (\ref{compact}) has an 
additional undesirable term.  Namely, the ordinary KK compactification 
of the $D$-dimensional Einstein gravity on $S^1$ by using the KK metric 
Ansatz given by Eq. (\ref{pertmet}) with ${\cal C}(z)=1$ leads to the 
following $(D-1)$-dimensional action:
\begin{equation}
S_{\rm KK}={1\over{2\kappa^2_{D-1}}}\int d^{D-1}x\sqrt{-g}h{\cal R}_g,
\label{ordkkact}
\end{equation}
where the scalar $h$ can be identified as a Brans-Dicke (BD) scalar 
of the BD theory \cite{bd1,bd2} with the BD parameter $\omega=0$, and the 
$(D-1)$-dimensional gravitational constant $\kappa^2_{D-1}$, in this 
case, is given in terms of the volume $V(S^1)$ of $S^1$ as $\kappa^2_{D-1}=
\kappa^2_D/V(S^1)$.  However, the effective $(D-1)$-dimensional action 
(\ref{compact}) in the domain wall background has an additional term, 
which is the potential term for the scalar $h$.  So, a $D$-dimensional 
gravitating object with non-trivial $h$ in a domain wall background does 
not effectively describe the corresponding $(D-1)$-dimensional configuration 
that would have been obtained through the ordinary KK compactification on 
$S^1$.  

Recently, some efforts \cite{iv,mk,mkl} have been made to understand 
the KK modes of graviton in the domain wall background associated with 
general perturbations around the non-dilatonic domain wall metric with 
non-trivial transverse perturbations.  It is observed there that the zero 
mode of transverse metric perturbation has to be zero, meaning that only  
solutions in the domain wall background with $h=1$ are allowed in the system 
with the action given by (\ref{ddwact}) plus (\ref{wvddwact}).  And it is 
further observed \cite{mkl} that the RS gauge (i.e., the metric perturbation 
of the form (\ref{metlngpurt}) with $h_{\mu\nu}$ satisfying the transverse 
traceless gauge condition) is classically stable.  In fact, the scalar 
potential in the effective action (\ref{compact}) has the minimum at $h=1$, 
implying that a solution with the metric (\ref{pertmet}) with $h=1$ is the 
classically preferred stable configuration.  So, we see that the result of 
Refs. \cite{iv,mk,mkl}, which is valid only perturbatively, continues to 
hold non-perturbatively, as well.  However, this does not mean that there 
does not exist solutions to the equations of motion of the action 
(\ref{ordkkact}) with non-trivial $h$, but it is just that the domain wall 
bulk background seems to prefer solutions with $h=1$.  An example of solution 
with a non-trivial $h$ in the system with the action (\ref{ordkkact}) is the 
$\omega=0$ case of the spherically symmetric solution with scalar hair
\footnote{Note, such a solution is not inconsistent with the Hawking's 
theorem \cite{nh} on black holes in the BD theory that the only 
stationary black holes in the BD theory are those of Einstein's 
general relativity, i.e., black hole solutions with the constant BD scalar.  
The reason is that the BD solution \cite{bd1,bd2} violates one of the 
conditions of the Hawking's theorem, namely, the weak energy condition.} 
constructed by BD \cite{bd1,bd2}.  On the other hand, as we will see in the 
following, a solution with non-trivial $h$ in the domain wall background 
is possible (or rather non-trivial $h$ is required in order for the 
effective theory in one lower dimensions to have vanishing cosmological 
constant term or no scalar potential term), if additional fields are added 
in the bulk action $S_{\rm bulk}$. 

It is interesting to note that the domain wall bulk background 
naturally induces a potential of the KK scalar $h$ (in the effective action 
(\ref{compact})), which was previously added to the KK action by hand in 
Ref. \cite{sol} in an attempt to remedy the problem of the KK theory that 
it corresponds to the BD theory with the BD parameter $\omega=0$, which is 
outside of the range of the constraint $\omega>500$ set by the solar system 
experiment \cite{rea,wil}.  Note, the solar system experiment does not 
constrain the allowed values of the BD parameter $\omega$, if the BD scalar 
has a potential with the minimum which allows the BD scalar to have a 
non-zero vacuum expectation value at low temperature.  A scalar potential 
was also added {\it ad hoc} with some justifications in the extended 
inflationary model \cite{ls}, which adopts a metric theory of gravity by 
introducing a scalar of the BD theory in an attempt to solve the ``graceful 
exit'' problem of the old inflation model \cite{kaz,gut} through slowing 
down of the inflationary expansion from exponential to power-law in time, 
in order to make the value of the BD scalar settle down at some large 
expectation value in the true-vacuum phase, i.e., after the inflationary 
phase \cite{wei,lsb}.  So, the noncompact compactification through the RS 
type domain wall appears to be more desirable than the ordinary KK 
compactification in this respect.  

Now, we consider the case when the domain wall bulk spacetime contains a 
charged $(p+1)$-brane, where one of the longitudinal directions of the 
brane is along the transverse direction of the domain wall, whose 
explicit solution is study in our previous work \cite{youm3}.  The total 
action $S$ for this case is give by the sum of the following bulk action:
\begin{equation}
S_{\rm bulk}={1\over{2\kappa^2_D}}\int d^Dx\sqrt{-G}\left[{\cal R}_G-
{4\over{D-2}}(\partial\phi)^2+e^{-2a\phi}\Lambda-{1\over{2\cdot(p+3)!}}
e^{2a_{p+1}\phi}F^2_{p+3}\right],
\label{bulkchbrn}
\end{equation}
the domain wall worldvolume action:
\begin{equation}
S_{\rm DW}=-\sigma_{\rm DW}\int d^{D-1}x\sqrt{-\gamma}e^{-a\phi},
\label{dwchbrn}
\end{equation}
and the following additional worldvolume action for the charged $(p+1)$-brane:
\begin{eqnarray}
S_{p+1}&=&-T_{p+1}\int d^{p+2}\xi\left[e^{-a_{p+1}\phi}\sqrt{-{\rm det}\,\,
\partial_aX^M\partial_bX^NG_{MN}}\right.
\cr
& &\ \ \ \ \ \ \left.+{\sqrt{\Delta_{p+1}}\over 2}{1\over{(p+2)!}}
\epsilon^{a_1...a_{p+2}}\partial_{a_1}X^{M_1}...\partial_{a_{p+2}}
X^{M_{p+2}}A_{M_1...M_{p+2}}\right],
\cr
\Delta_{p+1}&=&{{(D-2)a^2_{p+1}}\over 2}+{{2(p+2)(D-p-4)}\over{(D-2)}}.
\label{chbrnwvact}
\end{eqnarray}
The consistency of the equations of motion requires that the dilaton coupling 
parameters $a$ and $a_{p+1}$ satisfy the following constraint:
\begin{equation}
aa_{p+1}=-{{4(D-p-4)}\over{(D-2)^2}}.
\label{dilparacnst}
\end{equation}
Guided by the explicit solution presented in Ref. \cite{youm3}, we take 
Eq. (\ref{pertmet}) as the $D$-dimensional metric Ansatz and the following 
as the remaining fields Ans\" atze: 
\begin{equation}
e^{2\phi}={\cal C}^{{(D-2)^2a}\over 4}e^{2\tilde{\phi}},\ \ \ \ 
A_{p+2}={\cal C}^{{D-2}\over{2}}\tilde{A}_{p+2},
\label{ansatfld}
\end{equation}
where the tilded fields depend on the domain wall worldvolume coordinates 
$x^{\mu}$, only, and ${\cal C}$ is given by Eq. (\ref{ddcf}).  Substituting 
the above Ans\" atze for the fields into the total action $S$ and integrating 
over the extra space coordinate $z$, we obtain the following effective action:
\begin{eqnarray}
S&=&{1\over{2\kappa^2_D}}\int d^{D-1}xdz\sqrt{-g}\varpi^{2\over{\Delta+2}}
\left[h{\cal R}_g-{4\over{D-2}}h(\partial\tilde{\phi})^2-{1\over{2\cdot(p+2)!}}
h^{-1}e^{2a_{p+1}\tilde{\phi}}\tilde{F}^2_{p+2}\right.
\cr
& &\ \ \ \ \ \ \ \ \ \left.-8{{D-1}\over{D-2}}{Q\over\Delta}h^{-1}\left\{
\delta(z)-{{DQ}\over{4(D-1)}}\varpi^{-2}\right\}-{{2Q^2}\over\Delta}
\varpi^{-2}he^{-2a\tilde{\phi}}\right]
\cr
& &\ \ \ \ \ \ \ \ \ -{4\over{|\Delta|}}{Q\over{\kappa^2_D}}\int d^{D-1}x
\sqrt{-g}e^{-a\tilde{\phi}}+S_{p+1}
\cr
&=&{1\over{2\kappa^2_{D-1}}}\int d^{D-1}x\sqrt{-g}\left[h{\cal R}_g
-{4\over{D-2}}h(\partial\tilde{\phi})^2-{1\over{2\cdot(p+2)!}}h^{-1}
e^{2a_{p+1}\tilde{\phi}}\tilde{F}^2_{p+2}\right.
\cr
& &\ \ \ \ \ \ \ \ \ \left.+2{{(\Delta+4)Q^2}\over{\Delta^2}}(h^{-1}
+he^{-2a\tilde{\phi}}-2e^{-a\tilde{\phi}})\right]+S_{p+1},
\label{chbrndweffact}
\end{eqnarray}
where $\tilde{F}_{p+2}=d\tilde{A}_{p+1}$ with $(\tilde{A}_{p+1})_{\mu_1...
\mu_{p+1}}\equiv(\tilde{A}_{p+2})_{\mu_1...\mu_{p+1}z}$, 
$\kappa^2_{D-1}$ is given by Eq. (\ref{gravcnst}), and we let $\Delta<0$ 
and made use of the constraint (\ref{dilparacnst}) in the kinetic term for 
the form potential.   Note, the explicit expressions for $h$ and 
$e^{\tilde{\phi}}$ are given in terms of the harmonic function $H_{p+1}$ 
for the $D$-dimensional $(p+1)$-brane as \cite{youm3}:
\begin{equation}
h=H^{-{{2(D-p-4)}\over{(D-2)\Delta_{p+1}}}}_{p+1},\ \ \ \ \ \ \ 
e^{\tilde{\phi}}=H^{{(D-2)a_{p+1}}\over{2\Delta_{p+1}}}_{p+1}.
\label{hphiexprtn}
\end{equation}
So, the last line of the $(D-1)$-dimensional action in Eq. 
(\ref{chbrndweffact}) becomes zero.  The $(D-1)$-dimensional effective 
action in Eq. (\ref{chbrndweffact}), therefore, becomes of the form of the 
bulk action for the dilatonic $p$-brane in $D-1$ dimensions, obtained from 
the bulk action for the $D$-dimensional dilatonic $(p+1)$-brane through the 
ordinary KK compactification on $S^1$ along one of its longitudinal 
directions.  Next, we show that the effective action for the worldvolume 
action $S_{p+1}$ for the dilatonic $(p+1)$-brane in the bulk of 
$D$-dimensional domain wall has the form of the worldvolume action for the 
dilatonic $p$-brane in $D-1$ dimensions.  In the static gauge with constant 
transverse (to the $(p+1)$-brane) target space coordinates, the $(p+1)$-brane 
worldvolume action (\ref{chbrnwvact}) takes the following form:
\begin{equation}
S_{p+1}=-T_{p+1}\int d^{p+2}x\left[e^{-a_{p+1}\phi}\sqrt{-{\rm det}\,
G_{ab}}+{\sqrt{\Delta_{p+1}}\over 2}A_{tx_1...x_pz}\right],
\label{statggact}
\end{equation}
where $a,b=t,x_1,...,x_p,z$.  After substituting the Ans\" atze for the 
fields in the above and integrating over the extra space coordinate $z$, 
we obtain the following effective action:
\begin{eqnarray}
S_{p+1}&=&-T_{p+1}\int d^{p+1}xdz\,\varpi^{2\over{\Delta+2}}\left[e^{-a_{p+1}
\tilde{\phi}}h\sqrt{-{\rm det}\,g_{\tilde{a}\tilde{b}}}+{\sqrt{\Delta_{p+1}}
\over 2}\tilde{A}_{tx_1...x_pz}\right]
\cr
&=&-T_p\int d^{p+1}x\left[e^{-a_{p+1}\tilde{\phi}}h\sqrt{-{\rm det}\,
g_{\tilde{a}\tilde{b}}}+{\sqrt{\Delta_p}\over 2}\tilde{A}_{tx_1...x_p}
\right],
\label{pbrnact}
\end{eqnarray}
where $T_p\equiv -{{2\Delta}\over{(\Delta+4)Q}}T_{p+1}$ and $\tilde{a},
\tilde{b}=t,x_1,...,x_p$.  This is of the form of the worldvolume action for 
the $(D-1)$-dimensional $p$-brane obtained from the worldvolume action for 
the $D$-dimensional $(p+1)$-brane through the ordinary KK compactification
on $S^1$ along one of its longitudinal directions.  Therefore, the effective 
action for the $(p+1)$-brane in the bulk of $D$-dimensional dilatonic domain 
wall has the same form as the action for the $p$-brane in $D-1$ dimensions, 
obtained from the action for the $D$-dimensional $(p+1)$-brane through the 
ordinary KK compactification on $S^1$.

This result indicates that in the case where there are additional fields 
in the bulk action the gravity can be trapped within the domain wall 
even if the domain wall metric has a non-trivial perturbation along the 
transverse (to the domain wall) direction.  The zero mode of this transverse 
perturbation is identified as a scalar in one lower dimensions.  As we 
have seen from the dilatonic $(p+1)$-brane solution in the domain wall 
bulk, such zero mode of transverse metric perturbation conspires with the 
zero mode of the dilaton perturbation in such a way that the possible 
scalar potential term or the cosmological term in one lower dimensions 
is eliminated.  Thereby, a configuration in the domain wall bulk 
effectively describes the corresponding configuration in {\it an 
asymptotically flat spacetime} in one lower dimensions
\footnote{In Ref. \cite{youm3}, we argued that the dynamics of a test 
particle in the background of the charged $(p+1)$-brane in the bulk of the 
domain wall does not reproduce that of a test particle in the background 
of the corresponding $p$-brane in one lower dimensions because of the 
non-trivial dependence of the transverse (to the wall) component of the 
metric on the longitudinal coordinates of the wall.  This is due to the 
fact that the simple trick that was used in Refs. \cite{chr,youm3} to 
study the test particle dynamics (with non-trivial motion along the 
extra space direction) is not applicable in such case.  If we instead 
solve the full non-linear coupled geodesic equations $\textnormal{\it\" 
x}^{\mu}+\Gamma^{\mu}_{\nu\rho}\dot{x}^{\nu}\dot{x}^{\rho}=0$, 
which is rather a difficult task, it may be possible to reproduce the 
geodesic motion in one lower dimensions.}.  
One the other hand, the domain wall bulk spacetime is very restrictive 
about the possible gravitating configurations.  As pointed out in Refs. 
\cite{iv,mk,mkl} and also we have seen in the above, the gravitating 
configurations with a transverse (to the wall) metric component that 
depends on the longitudinal coordinates of the wall is not preferred, 
if there are no additional fields in the bulk action.  And only the dilatonic 
charged brane with the dilaton coupling parameter that satisfies the 
constraint (\ref{dilparacnst}) is allowed in the bulk of a dilatonic domain 
wall.  Because of this restriction, non-dilatonic domain wall bulk background 
cannot admit charged branes (with asymptotically flat spacetime in one lower 
dimensions) and the current RS type models (dilatonic and non-dilatonic) 
cannot admit, for example, the Reissner-Nordstrom black holes in one lower 
dimensions.  

We comment on an alternative realization of charged branes in the RS type 
models.   One may just want to regard charged branes as living within the 
domain wall.  Namely, one may regard the form fields that the charged branes 
couple to as being contained in the worldvolume action $S_{\rm DW}$ of the 
domain wall, not in the bulk action $S_{\rm bulk}$.  The combined bulk and 
worldvolume action has the following form:
\begin{eqnarray}
S&=&{1\over{2\kappa^2_D}}\int d^{D-1}xdz\sqrt{-G}\left[{\cal R}_G-{4\over{D-2}}
\partial_M\phi\partial^M\phi+e^{-2a\phi}\Lambda\right]
\cr
& &\ \ \ \ +\int d^{D-1}x\sqrt{-\gamma}\left[{\cal L}-e^{-a\phi}
\sigma_{\rm DW}\right],
\label{altact}
\end{eqnarray}
where the form fields that charged branes couple to are contained in the 
Lagrangian ${\cal L}$ on the domain wall worldvolume.  Then, by using the 
$D$-dimensional bulk metric Ansatz given by Eq. (\ref{pertmet}) with $h=1$, 
one can bring the action (\ref{altact}) to the following form of the effective 
$(D-1)$-dimensional action after integrating over the extra space coordinate 
$z$:
\begin{equation}
S={1\over{2\kappa^2_{D-1}}}\int d^{D-1}x\sqrt{-g}\left[{\cal R}_g-
{{\Delta+4}\over{2\Delta Q}}{\cal L}\right],
\label{alteffact}
\end{equation}
where the $(D-1)$-dimensional gravitational constant $\kappa_{D-1}$ is 
given by Eq. (\ref{gravcnst}).  An advantage of this approach is that 
there is no restriction on the possible gravitating configurations 
within the domain wall.  So, even the non-dilatonic domain wall of the 
original RS model \cite{ran1,ran2,ran3} can admit charged black hole 
solutions, including the Reissner-Nordstrom solution.  However, one of 
setbacks is that such description is inconsistent with the recent string 
theory view on charged black holes that enabled microscopic interpretation 
of the Bekenstein-Hawking entropy \cite{vs}.  Namely, if charged black 
holes are to be interpreted as being compactified from intersecting branes 
in ten or eleven dimensions, then the charged black holes have to be coupled 
to the fields in the bulk action $S_{\rm bulk}$.

\end{document}